# Towards an All-Silicon DV-QKD Transmitter Sourced by a Ge-on-Si Light Emitter

Florian Honz[(1)], Nemanja Vokić [(1)], Philip Walther[(2)], Hannes Hübel[(1)], and Bernhard Schrenk[(1)]

[(1)]Center for Digital Safety & Security, AIT Austrian Institute of Technology, 1210 Vienna, Austria, florian.honz@ait.ac.at
[(2)]University of Vienna, Faculty of Physics, Vienna Center for Quantum Science and Technology (VCQ), 1090 Vienna, Austria

*Abstract*—We investigate the behavior of a Ge-on-Si light source and demonstrate its feasibility for polarization-encoded discrete-variable quantum key distribution following the BB84 protocol, enabling a potential "all-silicon" QKD scheme which can operate well below the necessary QBER limit and successfully generate secret keys.

## I. Introduction

The rise of quantum computing threatens the security of our current cryptographic standards since it will be able to solve mathematical problems deemed too complex for our current, on classical bits based, computers exponentially faster. In this context quantum key distribution (QKD) can guarantee absolute security by capitalizing on the quantum properties of light. The development of QKD devices and systems has progressed far and earlier deployed systems have been successfully operated for many years. Nonetheless, QKD systems are still bulky and usually occupy at least one complete slot of a standard 19-inch server rack. As such they consume a considerable footprint in applications where space is precious, as it is the case for datacenters where high server densities of 10,000 and more require the seamless integration of communication equipment with storage and computing units [1]. Therefore, a simpler hardware implementation with reduced size and cost is sought after, something that photonic integration can provide [2-4]. A remaining challenge is the integration of the light source, which cannot be monolithically accomplished on silicon.

For this reason, we investigated the adoption of a Ge-on-Si light emitter as the source for a BB84 polarization-encoded QKD scheme, showing that QKD based on this simplified approach can generate a secret key and pave the way towards full monolithic integration of a QKD transmitter on silicon, as it would be required for the adoption of co-packaged QKD optics with compact footprint.

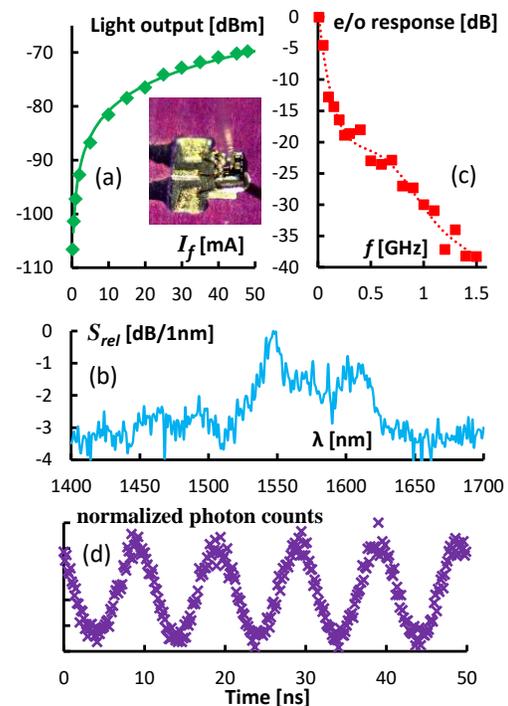

Fig. 1. (a) L-I characteristics, (b) emission spectrum, (c) electro-optic bandwidth of the Ge-on-Si emitter and (d) normalized photon counts for direct modulation of the emitter at 100 MHz.

## II. Ge-on-Si Light Emitter

For our experiment we used a die-level emitter which was wire-bonded to a PCB (Fig. 1(a)). The emitter is a forward-biased Ge-on-Si PIN junction which is vertically fiber-coupled. As can be seen in Fig. 1(a), the light-current (*L-I*) performance shows a LED-like behavior with an optical output that reaches up to -70 dBm, corresponding to an optical power of 100 pW, at a forward current of 46 mA. If we assume a symbol rate of 1 GHz for the QKD transmitter, which corresponds to -80 dBm at a mean photon number of $\mu_Q$ = 0.1 photons/symbol, there is an excess budget in light emission of ~10 dB. Therefore, this silicon emitter is strong enough to source QKD transmitters.

Figure 1(b) presents the emission spectrum of the PIN junction. While the peak emission wavelength of this sample is at 1547 nm, it can be clearly seen that the overall emission is LED-like and centered in the C+L band range, which is the main waveband range used in telecommunication applications with integrated QKD channels.

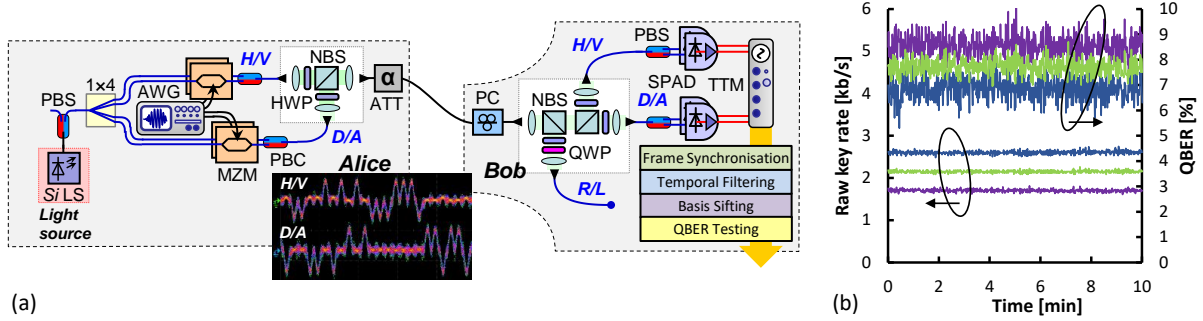

Fig. 2. (a) Experimental setup for evaluation of the proposed light source in QKD. (b) Raw-key rate and QBER.

Moreover, and to extend our initial findings [5], we investigated the electro-optic bandwidth of the Ge-on-Si emitter for the purpose of direct modulation (Fig. 1(c)). Since it eludes a conventional measurement with a vector network analyzer due to the operation at single-photon level, we instead employed a frequency-swept drive together with a single-photon avalanche photodetector (SPAD) and evaluated the magnitude of the contrast of this sinusoidal light modulation in the delay histogram of the time-tagged SPAD output. The direct response of the emitter to an exemplary modulation signal of 100 MHz is reported in Fig. 1(d). Despite 50Ω impedance matching of the PIN junction, we noticed a strong roll-off occurring at low frequencies (Fig. 1(c)). For this reason, we have resorted to external modulation for the state preparation when evaluating the applicability for QKD.

### III. EXPERIMENTAL SETUP. RESULTS AND DISCUSSION.

For the experimental evaluation of our emitter in a QKD scheme we used a polarization-encoded BB84 transmitter preparing the transmitted state in the four polarizations H/V/A/D (Fig. 2(a)). To accomplish this, we employed our Ge-on-Si light emitter as a continuous-wave source and used four Mach-Zehnder modulators (MZM) as optical gates, followed by polarization beam combiners (PBC) and half-wave plates (HWP) to create the target polarization states. Before transmission to Bob, the signal was attenuated to a mean photon number of $\mu_Q$ = 0.1 photons/symbol.

On Bob's side we first aligned the polarization of the incoming signal to our polarization analyzer with a manual polarization controller (PC). The polarization analyzer then transmits the incoming signal to one of four free-running InGaAs SPADs (efficiency 10%, dark count rate 560 cts/s), depending on the incoming polarization state. As can be seen from the inset in Fig. 2(a), there was no mixing between the different polarization states. The detection events of the SPADs were registered by a time-tagging module (TTM) to perform a real-time QBER estimation. This estimation included frame synchronization with Alice, using a pre-defined sequence, temporal filtering and basis sifting.

Figure 2(b) shows the measured raw-key rate and QBER for H (violet) and V (blue) as well as the average value (green). For a temporal filtering ratio of 50 %, the average QBER was 7.71 % at a raw key rate of 2.15 kb/s for the depicted H/V basis, with similar results for the A/D basis. This QBER, which is well below the threshold limit of 11% for which a secret key can be established [6], clearly indicates that an all-silicon QKD transmitter can be realized in the near future.

### IV. CONCLUSION

We experimentally characterized the applicability of a Ge-on-Si light emitter as a silicon-based light source for a future monolithic integrated QKD transmitter. The emission of the evaluated sample in the C+L band, together with its output power above the single-photon level is suitable for operation in a BB84 polarization-encoded QKD transmitter operating at a rate of 1 GHz. The transmitter yields a QKD performance well below the QBER limit at which a secret key can be established. This proves that components from standard silicon photonic platforms, without requiring process modifications, can be utilized as light sources for QKD applications, where a vast number of devices must be deployed at lowest possible cost and small footprint.

ACKNOWLEDGEMENT

This work was supported by the ERC under the EU Horizon-2020 programme (grant agreement No 804769) and by the Austrian FFG Research Promotion Agency and NextGeneration EU (grant agreement No FO999896209).

REFERENCES

[1] A. Singh et al., "Jupiter Rising: A Decade of Clos Topologies and Centralized Control in Google's Data center Network", ACM SIGCOMM Comp. Comm. Rev., vol. 45, no. 4 (2015).

[2] T. Paraiso et al., "A photonic integrated quantum secure communication system," Nature Phot., vol. 15, pp. 850-856 (2021).

[3] P. Sibso et al., "Chip-based quantum key distribution," Nature Comm., vol. 8, p. 13984 (2017).

[4] M. Avesani et al., "Full daylight quantum-key-distribution at 1550 nm enabled by integrated silicon photonics," npj Quant. Inf., vol. 7, p. 93 (2021).

[5] F. Honz et al., "Polarization-Encoded BB84 QKD Transmitter Sourced by a SiGe Light Emitter", Proc. Opt. Fiber Comm. Conf. (OFC), San Diego, United States, Mar. 2023, M1I.5.

[6] S. Pirandola et al., "Advances in quantum cryptography," Adv. Opt. Photon., vol. 12, pp. 1012-1236 (2020).